\documentclass{article}
\usepackage{spconf,amsmath,graphicx,hyperref}

\usepackage{algorithm, algorithmic, amssymb, booktabs, multicol, multirow}

\title{S$^2$Voice: Style-Aware Autoregressive Modeling with Enhanced Conditioning for Singing Style Conversion}

\name{Ziqian Wang$^{1}$, Xianjun Xia$^{2}$, Chuanzeng Huang$^{2}$, Lei Xie$^{1\ast}$\thanks{$\ast$  Corresponding author.}}
\address{
    $^{1}$ Audio, Speech and Language Processing Group (ASLP@NPU), School of Software, \\ Northwestern Polytechnical University, Xi'an, China \\
    $^{2}$ Bytedance, China
}
\begin{document}
%
\maketitle
\begin{abstract}
We present S$^2$Voice, the winning system of the Singing Voice Conversion Challenge (SVCC) 2025 for both the in-domain and zero-shot singing style conversion tracks. Built on the strong two-stage Vevo baseline, S$^2$Voice advances style control and robustness through several contributions. First, we integrate style embeddings into the autoregressive large language model (AR LLM) via a FiLM-style layer-norm conditioning and a style-aware cross-attention for enhanced fine-grained style modeling. Second, we introduce a global speaker embedding into the flow-matching transformer to improve timbre similarity. Third, we curate a large, high-quality singing corpus via an automated pipeline for web harvesting, vocal separation, and transcript refinement. Finally, we employ a multi-stage training strategy combining supervised fine-tuning (SFT) and direct preference optimization (DPO). Subjective listening tests confirm our system's superior performance: leading in style similarity and singer similarity for Task 1, and across naturalness, style similarity, and singer similarity for Task 2. Ablation studies demonstrate the effectiveness of our contributions in enhancing style fidelity, timbre preservation, and generalization. Audio samples are available~\footnote{https://honee-w.github.io/SVC-Challenge-Demo/}.
\end{abstract}
\begin{keywords}
singing style conversion, singing voice conversion, singing voice synthesis, SVCC 2025
\end{keywords}

\section{Introduction}
Voice conversion (VC) aims to modify a source speech waveform so that it sounds as if produced by a target speaker while preserving the original linguistic content. The Voice Conversion Challenge (VCC) series~\cite{toda2016voice, lorenzo2018voice, zhao2020voice} has fostered reproducible comparisons across competing systems by providing common datasets, evaluation protocols, and baseline implementations. Recently, interest has expanded from speech to singing: singing voice conversion (SVC) raises unique modeling and evaluation challenges due to wider pitch ranges, expressive timing, and diverse vocal techniques~\cite{liu2021diffsvc, jayashankar2023self, huang2023singing}.

Singing style conversion (SSC)~\cite{violeta2025serenade, zhang2025vevo} is a closely related but distinct problem: instead of changing singer identity, SSC seeks to alter \emph{how} a phrase is performed (the singing style) while maintaining the linguistic content and the source singer's identity. Compared to standard VC and SVC, SSC is more challenging because multiple valid style realizations may exist for the same musical content, and style changes must respect musical constraints such as melody and timing while remaining perceptually pleasing. Existing approaches that separate high-level modeling from acoustic rendering, have made progress but still face three core issues: (i) incomplete disentanglement between style and timbre, causing style/timbre leakage; (ii) limited ability of autoregressive models to robustly capture fine-grained style attributes; and (iii) insufficient high-quality singing data and training strategies for stable, generalizable conversion.

To advance SVC research, the Singing Voice Conversion Challenge (SVCC) 2025~\cite{violeta2025singing}~\footnote{https://www.vc-challenge.org/} defines two tasks: In-domain SSC (task~1), where the source singer appears in the training data and a same-singer style reference is available; and zero-shot SSC (task~2), where the source singer is unseen during training and must be converted using a style reference from a different singer. The zero-shot track is particularly demanding, making it an effective benchmark for style modeling and generalization.

In this work, we build upon the two-stage strong Vevo~\cite{zhang2025vevo} baseline and propose S$^2$Voice (\textbf{S}tyle-aware \textbf{S}inging \textbf{Voice}) to address the above issues. S$^2$Voice improves style control and timbre preservation through the following contributions: 
\begin{itemize}
  \item We integrate style embeddings into the AR LLM via FiLM-style~\cite{perez2018film} layer normalization and style-aware cross-attention to improve fine-grained style control.
  \item  We introduce a speaker-conditioned acoustic decoder that leverages pre-trained speaker embeddings to explicitly guide timbre preservation in the flow-matching transformer to enhance timbre similarity.
  \item We curate a large, high-quality singing corpus via an automated pipeline and employ a multi-stage training strategy (SFT + DPO~\cite{rafailov2023direct}) that together improves perceptual quality, stability, and zero-shot generalization.
\end{itemize}

Specifically, our submissions rank first in both SVCC 2025 tracks:  achieving the highest style similarity and singer similarity in task~1, and leading across naturalness, style similarity, and singer similarity in task~2. Through ablation studies, we show these components provide complementary gains for style fidelity, timbre preservation, and generation stability.

\section{Related Work}
\subsection{Disentangled Speech Representation}
A core challenge in voice and singing conversion is separating linguistic content from speaker timbre and expressive style. Recent systems rely on self-supervised learning (SSL) features~\cite{van2017neural} (e.g., HuBERT~\cite{hsu2021hubert}, Wav2Vec~\cite{baevski2020wav2vec}) and various discretization schemes to create compact, manipulable content representations that are easier to control. Vector-quantized encoders and clustering of SSL features (VQ-VAE~\cite{van2017neural}, k-means~\cite{ahmed2020k}) are widely used to produce discrete units that separate low-level acoustic information while retaining linguistic content~\cite{zhou2024zeroshotsingvoiceconversion}. 

However, the choice of tokenization strongly affects the trade-off between content integrity and leakage of timbre or style. Purely token-based pipelines can sometimes harm content accuracy or melodic fidelity in singing, while continuous embedding approaches risk carrying unwanted timbral cues~\cite{zhou2025simple}. To address this, prior work has explored hybrid solutions: augmenting discrete units with pitch or chroma features for melody preservation (e.g., TokSing~\cite{wu2024toksing}), applying duration reduction or run-length encoding to compress repeated tokens, and tuning codebook sizes to control the information passed to the high-level model~\cite{zhang2025vevo}. Other practical strategies include KNN replacement of SSL tokens from target speakers to reduce timbre mismatch~\cite{baas2023voiceconversionjustnearest}. Collectively, these studies emphasize that an explicit \emph{information bottleneck}—realized via quantization, pooling, or hierarchical tokenization—is essential for reliable disentanglement in VC/SVC, and that singing places additional demands (wider F0 range, vibrato, expressive timing) which token schemes should explicitly accommodate.


\subsection{Two-stage Frameworks}
Two-stage architectures that decouple high-level content/style modeling from low-level acoustic rendering have emerged as a robust paradigm for controllable synthesis and conversion~\cite{anastassiou2024seed, du2024cosyvoice}. In this paradigm, a high-level model (often autoregressive LLM-style) generates content or content–style representations, while a separate acoustic model maps these representations to spectrograms or waveforms and applies timbre conditioning. This separation simplifies conditioning and allows different architectures and losses to be applied at each stage.

Vevo~\cite{zhang2025vevo} is a prominent example of this idea: an autoregressive content–style transformer produces discrete content–style tokens, and a flow-matching transformer~\cite{lipman2022flow} renders spectrograms conditioned on a timbre reference. Vevo demonstrates that applying a VQ-based tokenizer to SSL features creates an \emph{information bottleneck}: by tuning the token vocabulary size, it balances content preservation against removal of timbre/style cues, thereby improving controllability and enabling strong zero-shot imitation. Complementary techniques used in the two-stage systems include duration reduction to shorten token sequences and classifier-free guidance or span masking in the acoustic flows to improve fidelity and robustness.

\section{Method}
\subsection{Overview}
We propose S$^2$Voice, a two-stage framework for controllable and zero-shot singing style conversion, building upon the Vevo architecture. In the first stage, we use an autoregressive language model to generate content-style tokens from input text or source audio, with style conditioning facilitated by a compact style encoder and modulated through FiLM-style layer normalization and style-aware cross-attention. In the second stage, we employ a flow-matching transformer to map content-style tokens to acoustic features, with timbre reference conditioning to maintain target speaker characteristics.



\subsection{Content--style Modeling with Style Conditioning}
The high-level module is an autoregressive Transformer (AR LLM)~\cite{touvron2023llama} that maps compact content tokens to content--style tokens conditioned on a style reference. Formally, given compressed content tokens \(\tilde{z}_c\) and a style reference \(r\), the AR model generates a sequence of content--style tokens
\begin{equation}
\tilde{z}_s = \mathrm{AR}(\tilde{z}_c \mid E_s),
\end{equation}
where \(E_s = \mathrm{Enc}_{\text{style}}(r)\) denotes the style embeddings from style encoder~\cite{chen2022wavlm, desplanques2020ecapa}.

\begin{figure}[t]
    \centering
    \includegraphics[width=0.9\linewidth]{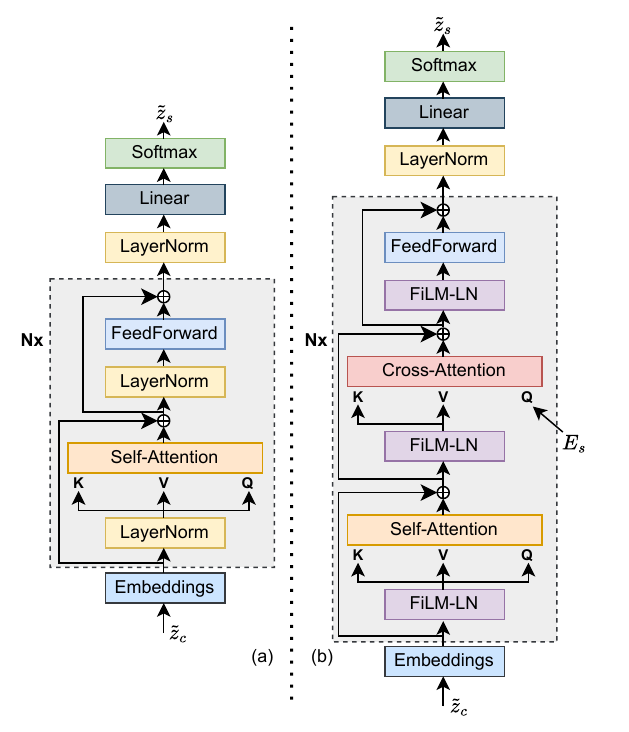}
  \caption{Autoregressive transformer block. (a) Original AR block with standard self-attention and feed-forward layers using conventional LayerNorm. (b) Modified AR block used in our AR-LLM: FiLM-style layer-norm modulation injects global style scale and shift (\(\gamma,\beta\)) produced by the style encoder, and a style-aware cross-attention lets local style queries attend to LLM context for fine-grained conditioning.}
    \label{fig:ar_block}
\end{figure}

Each Transformer block in the AR LLM is extended with two style-conditioning mechanisms: (i) FiLM-style modulation~\cite{perez2018film} applied to layer normalization, and (ii) a style-aware cross-attention block that fuses information from \(E_s\). A single Transformer block thus follows the sequence:
\begin{equation}
  \text{SelfAttn} \rightarrow \text{CrossAttn}_{\text{style}} \rightarrow \text{MLP},  
\end{equation}
with FiLM applied to the layer-norms inside the block, as shown in Fig.~\ref{fig:ar_block} (b).

Let \(h_i^\ell\) be the hidden vector at position \(i\) in layer \(\ell\). Denote the layer normalization as \(\mathrm{LN}(\cdot)\). We compute layer-wise scale and shift parameters from a projection of \(E_s\):
\begin{equation}
\gamma_\ell = W_\ell^\gamma \text{Proj}(E_s) + b_\ell^\gamma,\qquad
\beta_\ell = W_\ell^\beta \text{Proj}(E_s) + b_\ell^\beta. 
\end{equation}
The FiLM-modulated layer-norm output is
\begin{equation}
\mathrm{FiLM\text{-}LN}_\ell(h_i^\ell) = \big(1 + \gamma_\ell\big)\odot \mathrm{LN}(h_i^\ell) + \beta_\ell, 
\end{equation}
where \(\odot\) denotes element-wise multiplication. Applying FiLM in this way injects global style conditioning into each layer while keeping parameter overhead small.

To provide fine-grained, time-local style conditioning, we add a cross-attention module~\cite{vaswani2017attention} that allows the AR LLM to dynamically interact with the style embedding sequence \(E_s = [e_1^s,\dots,e_M^s]\). In our design, the cross-attention uses the style embeddings as \emph{queries} and the current LLM latents as \emph{keys} and \emph{values} so that style queries can collect matching content-context summaries:
\begin{equation}
Q = W_Q E_s,\qquad K = W_K H,\qquad V = W_V H,
\end{equation}
\begin{equation}
\mathrm{CrossAttn}(E_s, H) = \mathrm{softmax}\!\Big(\frac{QK^\top}{\sqrt{d}}\Big)V,
\end{equation}
where \(H\) denotes the LLM latent matrix at the relevant layer and \(d\) is the attention dimension. The cross-attention outputs are then projected and added into the transformer state. Intuitively, this lets each local style query attend to relevant content contexts and produce style-aware modulation signals.




\subsection{Acoustic Modeling with Speaker Conditioning}
In Vevo's two-stage pipeline, the acoustic decoder is conditioned by concatenating content--style tokens with tokens derived from a timbre reference and by using the timbre mel as a prompt to predict the target mel. Since the content--style tokenizer itself using summed Whisper features~\cite{radford2022robustspeechrecognitionlargescale} and chromagram contains style cues, and the timbre prompt may also carry stylistic traces, we observe that the acoustic stage can \emph{leak} style information from the timbre reference into the final waveform, which reduces perceived style conversion fidelity.

To mitigate this, we introduce an explicit, style-invariant global speaker condition into the acoustic flow model: a pre-trained speaker verification (SV) network~\cite{desplanques2020ecapa} produces a global speaker embedding \(s_g\) that represents speaker identity while being relatively insensitive to expressive style. The acoustic model then predicts mel-spectrograms conditioned on the AR-generated content--style token sequence \(\tilde{z}_s\) and the global speaker embedding \(s_g\), rather than relying solely on raw timbre prompts that may contain style cues.

Let \(\tilde{z}_s\) denote the content--style tokens output by the AR stage, and let \(x_{\text{spk}}\) be the waveform used to extract speaker identity. A pre-trained SV encoder \(\mathrm{SV}(\cdot)\) yields
\begin{equation}
s_g = \mathrm{SV}(x_{\text{spk}}) \in \mathbb{R}^d.
\end{equation}

We train the flow-matching~\cite{lipman2022flow} acoustic decoder \(p_\phi\) to model the conditional distribution
\begin{equation}
y \sim p_\phi\big(y \mid \tilde{z}_s,\; s_g\big),
\end{equation}
where \(y\) is the target mel-spectrogram. Concretely, the parameterized vector field of the continuous-time flow matcher is conditioned on both \(\tilde{z}_s\) and on \(s_g\). The flow-matching loss becomes
\begin{equation}
\label{eq:flow_speaker}
\mathcal{L}_{\mathrm{flow}} \;=\; \mathbb{E}_{y,\tau}\Big[\big\| v_\phi(y,\tau;\tilde{z}_s,s_g) - v^*(y,\tau)\big\|^2\Big],
\end{equation}
where \(v_\phi(\cdot)\) is the modeled vector field and \(v^*(\cdot)\) the target flow.

\subsection{Data Pipeline and Multi-stage Training}

We build an automated pipeline to assemble a large, high-quality singing corpus from web-harvested tracks and public datasets. Raw mixtures are processed with a pre-trained vocal separator~\cite{lu2024music} to extract vocal stems. For segments lacking transcripts, we run multiple ASR systems~\cite{radford2022robustspeechrecognitionlargescale, gao2022paraformer} and fuse their outputs: a token-level confidence score is computed from per-system confidences and inter-system agreement, and only high-confidence segments are kept. Transcripts are then refined with an LLM~\cite{yang2025qwen3} to correct recognition errors and normalize lyrical notation under prompts that favor musically plausible outputs. We apply automatic quality filtering using perceptual and signal metrics such as DNSMOS~\cite{reddy2022dnsmos}, framewise energy, pitch stability, and noise ratio. Finally, the retained data are deduplicated and balanced across styles, gender, and languages; the resulting curated corpus contains roughly 500 hours of high-quality singing vocals.

We adopt a multi-stage training recipe to tune the pre-trained models on singing-style conversion and to correct perceptual failures observed during inference:

We perform full-parameter supervised fine-tuning of both the AR LLM and the flow-matching acoustic model on the curated singing corpus. The AR LLM is fine-tuned with the next-token negative log-likelihood. The acoustic stage is fine-tuned with the flow-matching objective \(\mathcal{L}_{\mathrm{flow}}\).

To address failure modes observed during inference (e.g., early stopping, repetition, broken phrasing, jitter), we construct preference data from model outputs and apply a pairwise preference optimization stage~\cite{rafailov2023direct}. For each problematic input, we obtain a \emph{positive} example (ground-truth target) and one or more \emph{negative} examples (human-annotated model-generated bad-case outputs). Let \(x_{\text{pos}}\) denote the positive audio and \(x_{\text{neg}}\) a negative sample. We define a scoring function \(s_\theta(\cdot)\) derived from the model log-likelihood. The DPO-style pairwise preference loss we use is the standard logistic pairwise loss:
\begin{equation}\label{eq:dpo_loss}
\mathcal{L}_{\mathrm{pref}} \;=\; -\log\frac{\exp(s_\theta(x_{\text{pos}}))}{\exp(s_\theta(x_{\text{pos}}))+\exp(s_\theta(x_{\text{neg}}))}.
\end{equation}

\section{Experiments}

\subsection{Datasets and Evaluation Metrics}
We train and evaluate on two data sources. The first is the official SVCC 2025 training set provided by the challenge with a duration of approximately 70\, hours. The second is our curated large-scale singing corpus, the GTSinger corpus~\cite{zhang2024gtsinger} is \textbf{not} used in training as requested by the challenge.

For evaluation we conduct subjective listening tests and objective measures. Naturalness is reported as the mean opinion score (MOS). Style similarity and singer similarity are assessed with standard AB and XAB protocols. 

\subsection{Experimental Setup}
All models are initialized from pre-trained Vevo checkpoints and fine-tuned on the aforementioned datasets. Unless noted, model architecture and hyperparameters follow the Vevo defaults; we only modify learning rates for fine-tuning to adapt to singing data. Specifically, the learning rates for the AR LLM and flow-matching transformer are $2 \times 10^{-5}$ and $7 \times 10^{-6}$ during SFT, respectively. After SFT, we apply DPO only to update the AR LLM with a learning rate of \(1\times10^{-6}\).

\subsection{Experimental Results}
\label{sec:exp_results}
We evaluate our system on two tracks of SVCC 2025, comparing against the Vevo baseline and conducting ablations on Task 2. All scores are with a 95\% confidence score from the official listening test. ``GT''  denotes ground truth.

As shown in Table~\ref{tab:main_results}, S$^2$Voice achieves top-ranked performance, outperforming Vevo across all metrics. The largest gains are in \textit{Style Similarity}, validating our FiLM and cross-attention design. \textit{Singer Similarity} also sees strong improvement, confirming the efficacy of the global speaker embedding in improving timbre similarity.


\begin{table}[htbp]
\centering
\caption{Subjective evaluation results on SVCC 2025.}
\label{tab:main_results}
\resizebox{\linewidth}{!}{%
\begin{tabular}{l c ccc}
\toprule
\textbf{System} & \textbf{Task} & \textbf{Naturalness (MOS)} & \textbf{Style Sim. (\%)} & \textbf{Singer Sim. (\%)} \\
\midrule
GT       & 1 & 3.90 ± 0.15 & 79 ± 3 & 63 ± 4 \\
Vevo     & 1 & 3.10 ± 0.12 & 30 ± 5 & 42 ± 5 \\
\textbf{S$^2$Voice} & 1 & \textbf{3.30 ± 0.10} & \textbf{59 ± 4} & \textbf{57 ± 4} \\
\midrule
GT       & 2 & 4.10 ± 0.15 & 78 ± 3 & 60 ± 4 \\
Vevo     & 2 & 3.20 ± 0.12 & 32 ± 5 & 52 ± 5 \\
\textbf{S$^2$Voice} & 2 & \textbf{3.75 ± 0.11} & \textbf{70 ± 3} & \textbf{59 ± 4} \\
\bottomrule
\end{tabular}%
}
\end{table}


We further conduct an ablation study on Task~2, starting from SFT-only training on the curated corpus and progressively adding proposed components. As shown in Table~\ref{tab:ablation},  each component contributes positively: the curated data pipeline establishes a strong baseline; FiLM and style-aware cross-attention progressively enhance style modeling; and the global speaker embedding substantially improves singer similarity. Notably, while the final DPO stage leads to a marginal decrease in metrics, it reduces failure modes—such as repetitions, abrupt cuts, and jitter—that degrade listening quality. By optimizing for human preference, DPO trades a slight reduction in average metrics for fewer low-quality outliers, resulting in more stable and perceptually preferred outputs, particularly in out-of-domain scenarios.



\begin{table}[htbp]
\centering
\caption{Ablation study on SVCC 2025 Task 2 (Zero-shot).}
\label{tab:ablation}
\resizebox{\linewidth}{!}{%
\begin{tabular}{l|ccc}
\toprule
\textbf{Model Variant}       & \textbf{Naturalness (MOS)} & \textbf{Style Sim. (\%)} & \textbf{Singer Sim. (\%)} \\
\midrule
SFT Only                     & 3.50 ± 0.12                & 62 ± 4                   & 52 ± 5                    \\
+ FiLM                       & 3.62 ± 0.11                & 65 ± 4                   & 54 ± 4                    \\
+ Cross-Attention            & 3.68 ± 0.11                & 68 ± 3                   & 56 ± 4                    \\
+ Global Spk. Emb.            & \textbf{3.75 ± 0.11}       & \textbf{70 ± 3}          & \textbf{59 ± 4}           \\
+ DPO                       & 3.72 ± 0.11                & 69 ± 3                   & 58 ± 4                    \\
\bottomrule
\end{tabular}%
}
\end{table}





\section{Conclusion}
We present S$^2$Voice, our winning system for the SVCC 2025 challenge on singing style conversion. Building on the Vevo baseline, we advance style control and robustness through FiLM-based layer-norm modulation, style-aware cross-attention, and a global speaker embedding to improve timbre similarity. A large-scale curated singing corpus and a two-stage training strategy with SFT and DPO further enhance quality and stability. S$^2$Voice ranks first in both in-domain and zero-shot tracks, achieving state-of-the-art performance across naturalness, style similarity, and singer similarity.

\vfill\pagebreak




\footnotesize
\bibliographystyle{IEEEbib}
\bibliography{strings,refs}

\end{document}